# Comparing Single Molecule Tracking and correlative approaches: an application to the datasets recently presented in Nature Methods by Chenuard et al.

Carmine di Rienzo[1,2] and Paolo Annibale[3]

[1] NEST, Scuola Normale Superiore and Istituto Nanoscienze-CNR, Piazza San Silvestro 12, 56127 Pisa, Italy (carmine.dirienzo@sns.it)

[2] Center for Nanotechnology Innovation @NEST, Istituto Italiano di Tecnologia, Piazza San Silvestro 12, 56127 Pisa, Italy

[3] Department of Pharmacology, University of Wuerzburg, Wuerzburg, Germany (paolo.annibale@alumni.epfl.ch).

Recent efforts to survey the numerous softwares available to perform single molecule tracking (SMT) highlighted a significant dependence of the outcomes on the specific method used, and the limitation encountered by most techniques to capture fast movements in a crowded environment [1]. Other approaches to identify the mode and rapidity of motion of fluorescently labeled biomolecules, that do not relay on the localization and linking of the images of isolated single molecules are, however, available [2, 3].

In particular, the molecular Mean Square Displacement (MSD) can be easily measured without identifying individual molecules positions by quantifying the enlargement in time of the Spatio-Temporal Image Correlation Function (STICS), the so called iMSD approach [2]. Alternatively, the iMSD can be applied to the spatio-temporal correlation of localization datasets, building on the improved spatial resolution obtained by the molecular localization process [3, 4]. We set to apply each of these methods to the benchmark dataset introduced by Chenouard et al. [1]. We chose the scenario where a "receptor" molecule undergoes diffusion in between intervals of directed motion, associated to "transport" along the cytoskeleton. Three exemplary levels of signal-to-noise (SNR) and particle concentrations were investigated as shown in Fig. 1.

The conventional SMT analysis fails to quantitate accurately the presence of a directed motion (Fig. 1 a-d-g), due to the difficulty of correctly relinking the localizations of the fast moving molecules. This results in a sizable underestimation of the MSD, particularly at longer delays as shown by Chenuard et al. [1]. On the other hand, the iMSD (see supplementary information for further details) allows an accurate recovery of the expected MSD (Fig. 1 b-e-h). When this approach is applied to the localization datasets, the increased spatial resolution enables to resolve in the correlation function a peak indicating the presence of a directed motion (arrows in Fig. 1 c-f-i, Supplementary Fig. 2 and the supplementary note 1), enabling to quantify the simulated speed of 4-5 pixels/frames (see Supplementary Fig. 2).

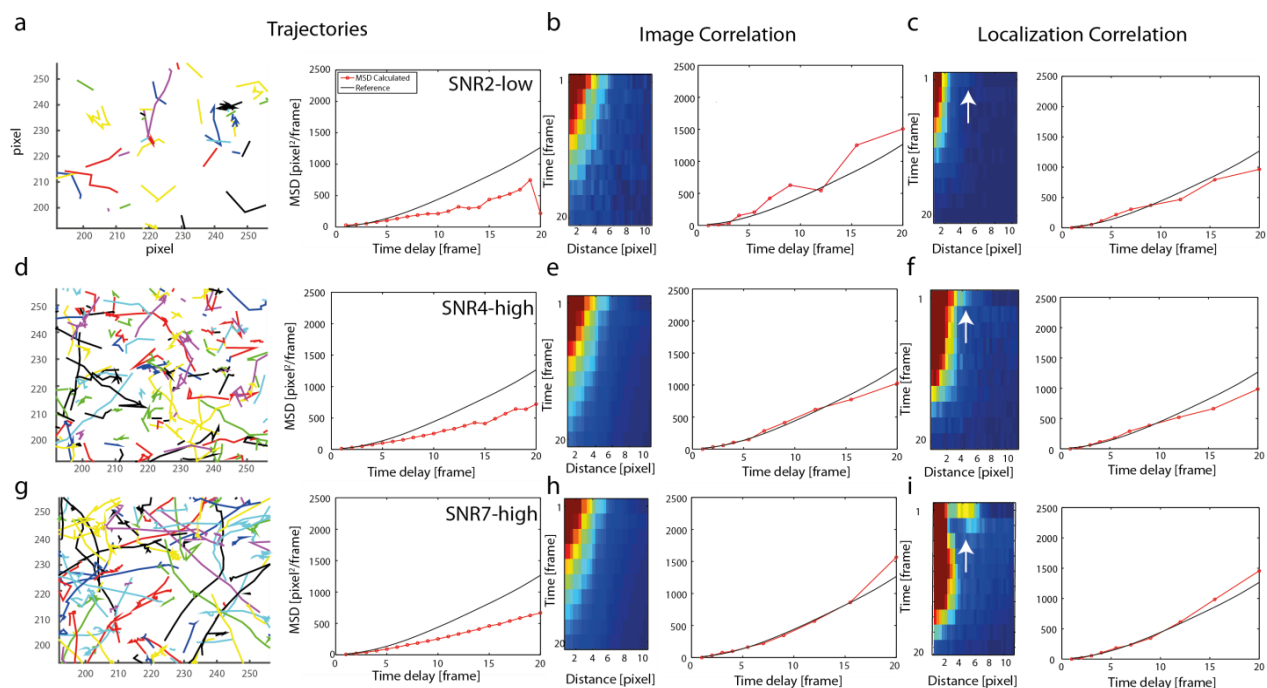

**Figure 1** Recovering the characteristic of the motion by: (a,d,g) classical SMT approach, (b,e,h) by the iMSD approach to STICS function of the image series, and (c,f,i) by the iMSD approach to the spatial-temporal correlation function of localization dataset. Case 1: Low signal to noise ratio (SNR 2) and low molecular expression. a) Individual molecules are localized frame by frame and the trajectories are linked left: linked trajectories superposed to molecular localizations. Right: the MSD plot from the experimental data is displayed on top of the MSD calculated from the "ground" values provided by the Chenouard et al. [1]. b) The molecular motion is extracted from the STICS function of the image series (left). The resulting iMSD plots are displayed (right). c) Molecular positions are first identified, frame by frame, using the localization algorith, and then binary images generated from these measurements are analyzed using a correlative approach. The STICS function extracted using this approach is displayed (left). The peak due to the fast directed motion of the receptors can now be resolved in the correlation carpet displayed in the figure (arrow). The resulting MSD curves are displayed in the right panel. The same legend applies to Case 2 (SNR 4 and high molecular expression represented in panels d-f) and Case 3 (SNR 7 high molecular expression represented in panels g-i).

As a further example, Supplementary Fig. 3 reports the analysis of fluorescently labeled β2 adrenergic receptor measured by single particle tracking-Photoactivated Localization Microscopy (spt-PALM)[5]. Blinking of the fluorophore, mEos2 in this case [6], prevents to obtain trajectories longer than few hundreds of milliseconds (inset in Supplementary Fig. 3a). On the other hand, the correlation function of the localized molecules clearly shows a peak at 0.8 μm after about 2 s, indicating the presence of active regulation of receptor motion (Supplementary Fig. 3b).

In summary, this direct comparison shows that correlative imaging analysis approaches complement effectively current SMT methods in circumstances when, due to either the density of the sample, the low signal to noise ratio or molecular blinking, trajectory linking does not allow to capture long-range or fast motion.

# Supplementary Figures

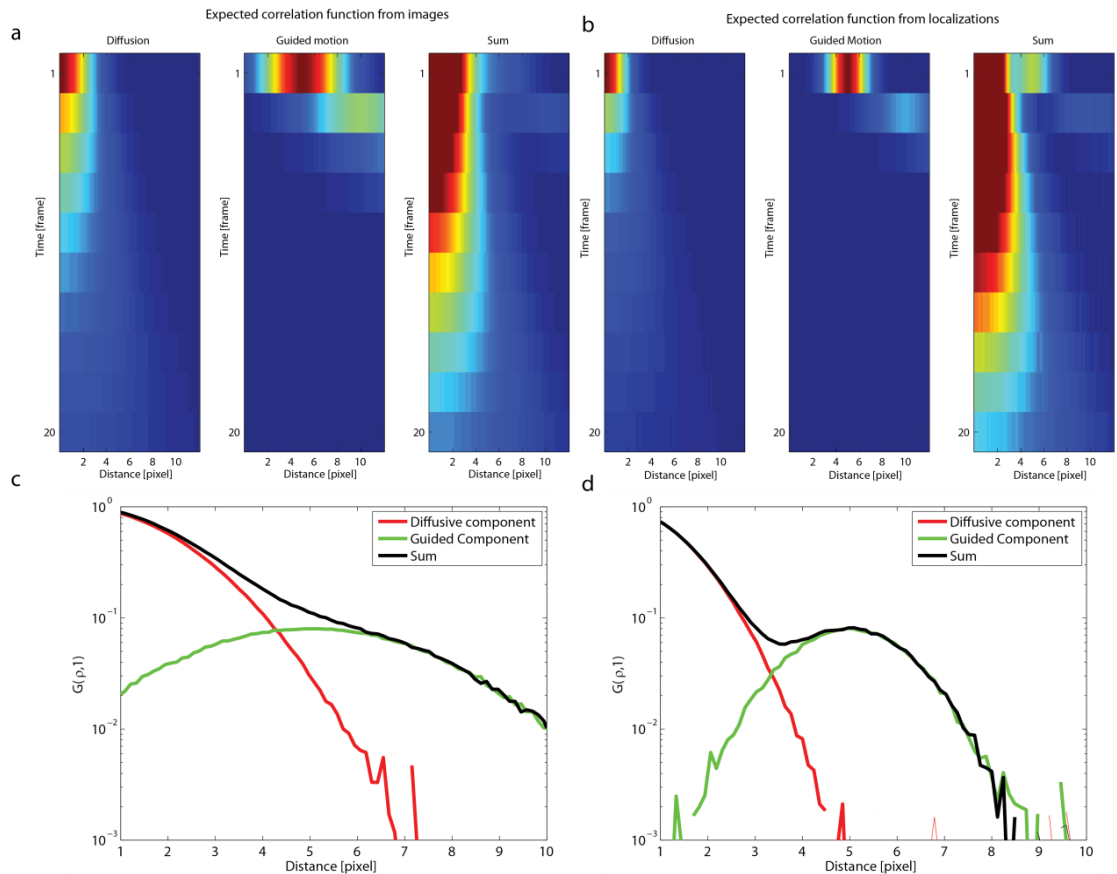

**Supplementary Figure 1: Theoretical Spatial-temporal correlations illustrating the combination of diffusive and directed motion.** $G(\boldsymbol{\rho},\boldsymbol{\tau})$ a) Correlation functions for a population of molecules partly diffusing at 1 pix$^2$/frame (left: diffusion) and partly subjected to a directed motion of 5 pix/frame (center: Flux), and the combination of the two types of motion (right: Sum). Spatial resolution is that of the image, namely $w_0$=2 pixels. b) The same $G(\boldsymbol{\rho},\boldsymbol{\tau})$ as in a, with an increased spatial resolution arising from the localization of the individual molecules, $w_0$=0.5 pixels. c) $G(\boldsymbol{\rho},\boldsymbol{\tau})$ in a at a delay of 1 frame: due to the limited spatial resolution (2 pixels), the sum of the diffusive component and of the peak due to the directed motion masks the fingerprint of the directional motion. d) This is not the case with an enhanced spatial resolution of 0.5 pixels, where the peak is clearly resolvable also in the Sum image.

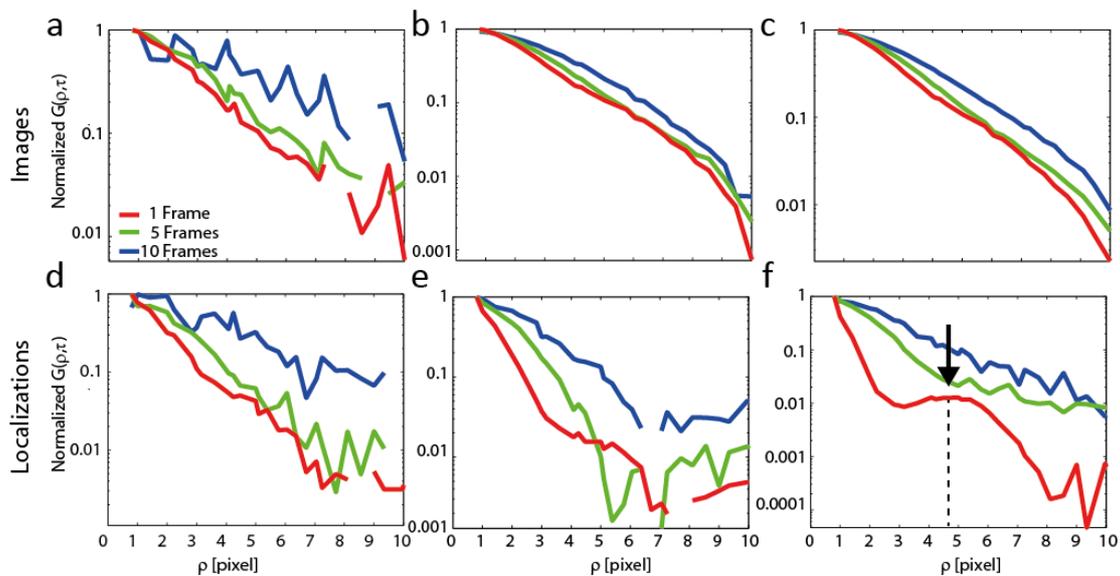

**Supplementary Figure 2 Correlation curves extracted from the images (top) and the localizations (bottom) for the three cases reported in Fig. 1.** Panels a) and d) represent Case 1. Panels b) and e) represent Case 2. Panels c) and f) represents Case 3. Three correlation profiles are shown, corresponding to a delay of 1, 5 and 10 frames. While the profiles from the images display only a widening at larger frame lags due to the effect of diffusion, the correlation functions calculated on the localizations allow to measure the presence of a peak (black arrow), which reflect the presence of a discrete step (~5 pixel) directional motion in the simulated dataset.

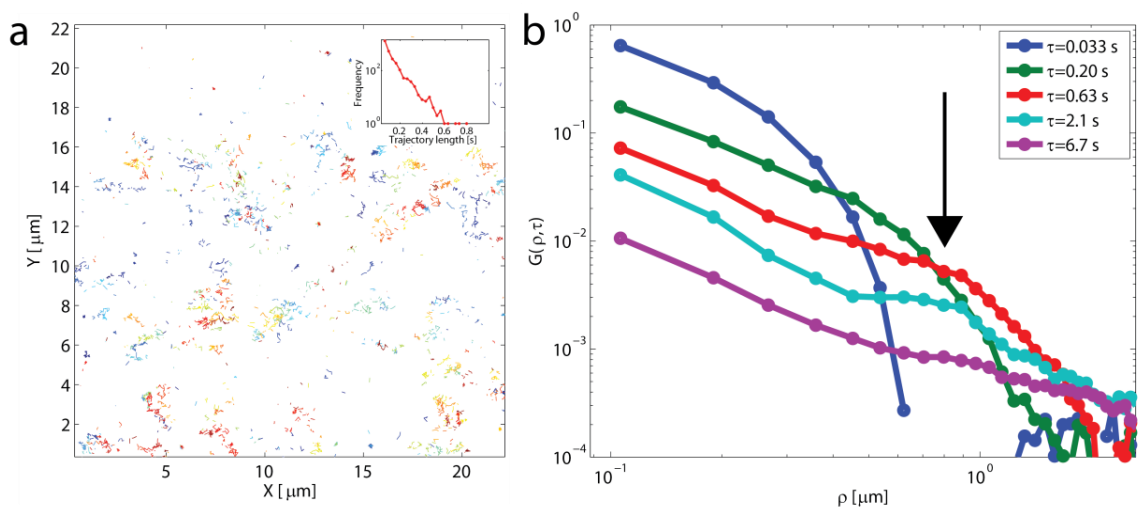

**Supplementary Figure 3 Correlative analysis on a single molecule experimental dataset.** a) Trajectories of the receptor ß2-Adrenergic Receptor fused to the photoactivatable fluorescent proteins mEos2, imaged in TIRF on the plasma membrane of a Cardiomyocyte-like cells H9C2. b) $G(\boldsymbol{\rho},\tau)$ measured on the localization dataset is shown for time lags increasing up to 6.7 s, analogous to those illustrated in Supplementary Fig. 2. In analogy to the simulated datasets, the clear peak in $G(\boldsymbol{\rho},\tau)$ indicate the presence of guided motion of the receptor. (Data are those published in Scarselli et al., *Journal of Biological Chemistry*, **2012**[7]).

# Supplementary Information

## *Simulated Dataset:*

The simulated datasets used in this work were recovered from the repository provided by Chenuard et al [1]. According to the authors, the chosen simulation type, i.e. Scenario 3 ("Receptors"), contains simulated particles switching between Brownian and directed motion models (with fixed probability), imaged in 2D+time using a confocal microscope (single plane mode). This type of dynamics may be observed with, for example, various types of receptor and motor proteins. The dynamics model of the particles was implemented as follows. For every particle, going from the current time point to the next, Brownian motion was simulated by sampling from a normal (Gaussian) distribution centered at the current position and with standard deviation 0.6 pixels. Directed motion was simulated as near-constant velocity with small random accelerations to allow deviations from a straight path. Concretely, for every particle, the position in the next time point was computed as the current position plus the displacement dictated by the current velocity vector.

During simulation the velocity was kept in (clipped to) the range from 2 to 6 pixels per frame.

The datasets were analyzed using custom-written Matlab routines to extract Spatial-Temporal Correlation Functions from the raw images as well as from the images results of the localization process according to Eq. 1. In the latter case, the intensity is considered as 1 for **r** corresponding to localization and 0 elsewhere.

## *Data Acquisition*

The live cell imaging data that were analyzed were among those published in Scarselli et al. 2012[7]. Experimental setup and data aquisition parameters are described therein. Briefly, the photoactivatable fluorescent protein mEos2 was fused to the G Protein-Coupled Receptor protein ß2 (ß2-mEos2), allowing to track a sparse subset of the molecules as they diffuse on the plasma membrane by subsequent photoactivation and photobleaching cycles. Live cell imaging was performed at 50 Hz in frame transfer mode. Cardiomyocites-like cells (H9C2) were imaged in TIRF mode under low 561 nm excitation at Room Temperature under a 100 X objective and an additional 2x magnification lens yielding a pixel size of 80 nm.

## *SMT:*

SMT was performed using the MOSAIC ImageJ Plugin as shown by Sbalzarini et al [8]. In the case of ß2-mEos2, trajectories longer than 5 frames were reconstructed using a linking range of 60ms (3 frames) and a displacement of 0.24 μm (~2 pixels). Mean Square Displacement was calculated at each time value as the average of the MSDs from all the measured trajectories.

## *iMSD analysis:*

The STICS correlation function was measured as:

$$G(\boldsymbol{\rho}, \tau) = \frac{\langle I(\boldsymbol{r},t)I(\boldsymbol{r}+\boldsymbol{\rho}, t+\tau)\rangle}{\langle I(\boldsymbol{r},t)\rangle^2} - 1, \qquad (1)$$

where $I(\boldsymbol{r},t)$ represents the intensity measured in the position $\boldsymbol{r}$ and at time t and $\langle ... \rangle$ represents the average of space and time. In the classical iMSD approach, $G(\boldsymbol{\rho}, \tau)$ is fitted to a Gaussian function defined as:

$$G(\boldsymbol{\rho}, \tau) = g(\tau) exp\left(-\frac{|\rho|^2}{\sigma(\tau)^2}\right) + g_\infty(\tau), \qquad (2)$$

where $g(\tau)$ represents the amplitude of the correlation function, $g_\infty(\tau)$ an offset usually necessary to take into account for mis-normalization of the correlation function [9, 10], and $\sigma(\tau)^2$ represents the iMSD and contains the contribution of particle motion and instrumental point spread function. In particular, it can be generically expressed as $\sigma(\tau)^2 = MSD(\tau) + \sigma_0^2$ where $\sigma_0^2$ is the PSF waist. However, independently on the particle motion, in diluted conditions, the spatio-temporal correlation function can be expressed as the spatial convolution of $P(\boldsymbol{\rho}, \tau)$, the probability to find a particle at a distance $\boldsymbol{\rho}$ after a time $\tau$, and the autocorrelation function of the PSF ($W(\boldsymbol{\rho})$) [2]:

$$G(\boldsymbol{\rho}, \tau) = \int P(\boldsymbol{\rho}, \tau) W(\boldsymbol{\rho} - \boldsymbol{r}) d\boldsymbol{r} + g_\infty(\tau). \qquad (3)$$

Moreover, it should be noted that the MSD, can be related to $P(\boldsymbol{\rho}, \tau)$ in a 2 dimensional motion by the relation:

$$MSD(\tau) = \int |\boldsymbol{\rho}|^2 P(\boldsymbol{\rho}, \tau) d\boldsymbol{\rho}, \qquad (4)$$

that represents definition of second moment of the probability distribution function $P(\boldsymbol{\rho}, \tau)$. As a consequence:

$$\sigma(\tau)^2 = \frac{\int |\boldsymbol{\rho}|^2 (G(\boldsymbol{\rho}, \tau) - g_\infty(\tau)) d\boldsymbol{\rho}}{\int (G(\boldsymbol{\rho}, \tau) - g_\infty(\tau)) d\boldsymbol{\rho}} \qquad (5)$$

Thus, in order to quantify $\sigma(\tau)^2$ could be in principle obtained by integrating $G(\boldsymbol{\rho}, \tau)$. In order to adress this issue, we fitted to a proper chose combination of Gaussian function the measured $G(\boldsymbol{\rho}, \tau)$. Three Gaussian functions were enough to describe the correlation function in all cases. The fitted $G(\boldsymbol{\rho}, \tau)$ is than substituted in Eq. 5 in order to calculate $\sigma(\tau)^2$.

## *Supplementary Note 1: the expected correlation function of simple models of motion*

In the case of 2D simple diffusion, the STICS correlation function can be expressed as:

$$G(\boldsymbol{\rho}, \tau) \propto \frac{exp\left(-\frac{|\boldsymbol{\rho}|^2}{\sigma_0^2 + 4D\tau}\right)}{\sigma_0^2 + 4D\tau}, \qquad (6)$$

where D is the diffusion coefficient. The STICS correlation function calculated by Eq. 6 is presented in Supplementary Fig. 1a and b (left panels) for D=1 pixel²/frame and two different $\sigma_0$, 2 and 0.5 pixel, as examples of correlation function obtained from images and sub-resolution localization respectively.

On the other hand, as already shown before[10], the STICS correlation function in the case of active transport can be expressed as:

$$G(\boldsymbol{\rho}, \tau) \propto exp\left(-\frac{|\boldsymbol{\rho} - \boldsymbol{v}\tau|^2}{\sigma_0^2}\right), \qquad (7)$$

where $\boldsymbol{v}$ is the velocity of active transport. In the simulated dataset of Chenouard et al. [1] the velocity is not constant. In order to take into account the variance introduce by the fluctuation of velocity in time, we approximated the correlation function as:

$$G(\boldsymbol{\rho}, \tau) \approx \frac{exp\left(-\frac{|\boldsymbol{\rho} - \boldsymbol{v}_m \tau|^2}{\sigma_0^2 + \sigma_v^2 \tau}\right)}{\sigma_0^2 + \sigma_v^2 \tau}, \qquad (8)$$

where $\boldsymbol{v}_m$ is the average speed and $\sigma_v^2$ represents the variance as calculated at 1 frame delay.

The STICS correlation function calculated by Eq. 8 is presented in Supplementary Fig. 1a and b (middle panels) for the same $\sigma_0$ mentioned before where $\boldsymbol{v}_m$ was set to 5 pixel/frame and $\sigma_v$ to 2 pixel/frame. Please note that differently from normal diffusion, in the case of active transport the peak of correlation function is not at $\boldsymbol{\rho} = \boldsymbol{0}$ but it shifts as a function of time according to the speed of active transport.

The correlation function obtained by Eq. 6 and Eq. 8 were summed in order to approximate the expected correlation function obtained by the combination of simple diffusion and directed motion. The results in shown in the right panels of Supplementary Fig. 1a and b.

The profile of the STICS correlation function obtained for $\tau = 1$ frame is reported in Supplementary Fig. 1c and d in order to highlight the contribution of simple diffusion (red lines) and active transport (green lines) to the correlation function (black lines). In the case of $\sigma_0 = 2$ the contribution of active transport cannot be easily recognized due to the poor spatial resolution and it appears as a tail of the correlation function. On the contrary, for $\sigma_0 = 0.5$ a clear peak appears in the correlation function profile enabling to recognize the presence of active transport by simple visual inspection. Please note that the position and the width of the peak enable to estimate the transport speed and its dispersion respectively.

A similar behavior is observed in the STICS correlation function calculated on the sample dataset provided by Chenouard et al. [1] (Supplementary Fig. 2). In particular, for high SNR the STICS correlation function calculated on localization dataset displays a clear peach for $\tau = 1$ indicating the presence of an active transport (red line in Supplementary Fig. 2f). Moreover, the position of such peak confirms that an average velocity of 4-5 pixels/frames was used when simulating the directed part of the particle motion. Please note that the spatial resolution enhancement gained by the particle localization step allows us obtaining a much better resolved correlation function for the system, in fact such information cannot be obtained by the correlation function obtained from images in similar conditions (red line in Supplementary Fig. 2c).